\documentclass[pra,aps,showpacs,twocolumn]{revtex4}
\usepackage{amsfonts}
\usepackage{amsmath}
\usepackage{graphicx}
\usepackage{dcolumn}
\usepackage{bm}

\begin{document}
\title{Nonlinear dynamics of a cigar-shaped Bose-Einstein condensate coupled with a single cavity mode}
%\author{J.~M. Zhang$^{1,2}$, F.~C. Cui$^1$, Y. Zhao$^2$, D.~L. Zhou$^1$, and W.~M. Liu$^1$}
%\affiliation{$^1$Beijing National Laboratory for Condensed Matter
%Physics, Institute of Physics, Chinese Academy of Sciences, Beijing
%100080, China\\
%$^2 $School of Materials Science and Engineering, Nanyang
%Technological University, Singapore 639798, Singapore}

\author{J.~M. Zhang, F.~C. Cui, D.~L. Zhou, and W.~M. Liu}
\affiliation{Beijing National Laboratory for Condensed Matter
Physics, Institute of Physics, Chinese Academy of Sciences, Beijing
100080, China}
\begin{abstract}
We investigate the nonlinear dynamics of a combined system which is
composed of a cigar-shaped Bose-Einstein condensate and an optical
cavity. The two sides couple dispersively. This system is
characterized by its nonlinearity: after integrating out the freedom
of the cavity mode, the potential felt by the condensate depends on
the condensate itself. We develop a discrete-mode approximation for
the condensate. Based on this approximation, we map out the steady
configurations of the system. It is found that due to the
nonlinearity of the system, the nonlinear levels of the system can
fold up in some parameter regimes. That will lead to the breakdown
of adiabaticity. Analysis of the dynamical stability of the steady
states indicates that the same level structure also results in
optical bistability.
\end{abstract}
 \maketitle
\section{introduction}
Recently, a lot of investigations have been devoted to the
combination of ultracold atom physics and cavity quantum
electrodynamics. Experimentally, the efforts culminate in the
successful coupling of a Bose-Einstein condensate (BEC), that is, a
single matter-wave field mode, to a single cavity mode in a high
finesse optical cavity \cite{esslinger_N,colombe_N}. Though of a
very short history, this type of combined system has demonstrated
many interesting phenomena, such as vacuum Rabi splitting
\cite{esslinger_N,colombe_N}, $\sqrt{N}$ scaling of the atom-photon
coupling \cite{esslinger_N,colombe_N}, optical bistability
\cite{stamper-kurn_prl,chapman}, cavity enhanced superradiance of a
BEC \cite{courteille}, and most surprisingly, a map between the atom
ensemble-cavity system and the canonical opto-mechanical system is
found \cite{stamper-kurn_NP, esslinger_a}. Besides playing an active
role in the dynamics, the cavity can also be used to characterize
the properties of the matter-wave field. It has been experimentally
implemented to study the statistics of an atom laser by detecting
single atoms falling through a cavity \cite{esslinger_pra}. Based on
similar principles, a proposal to probe the superfluidity-Mott
insulator transition is brought up \cite{ritsh_NP}.

One of the most remarkable characteristics of the atom
ensemble-cavity system is its intrinsic nonlinearity. If we are only
interested in the mechanical motion of the atoms and thus confine
ourself to the large detuning limit, the atom-photon interaction is
of a dispersive nature. The atom-photon interaction provides a
potential (an optical lattice) to the atoms, and meanwhile shifts
the cavity mode frequency. By shifting the cavity mode frequency,
the atom ensemble can influence or even determine the potential
acting on itself. This entails a nonlinearity substantially
different than the usual atom-atom interactions. It is this
nonlinearity that lies at the heart of previous experimental and
theoretical works
\cite{esslinger_a,stamper-kurn_prl,stamper-kurn_NP,hemmerich,ritsch_pra_00,zhang},
in particular, the dispersive optical bistability
\cite{stamper-kurn_prl,hemmerich}.

Along this line, we consider the nonlinear dynamics of a
cigar-shaped Bose-Einstein condensate in an optical cavity in this
work. Experimentally, this has been achieved and investigated by
Esslinger \textsl{et al}. \cite{esslinger_a}, and in the weak
atom-atom interaction and weak excitation limit (or non-depletion
limit), it is found that the condensate can be described as a
harmonic oscillator and the condensate-cavity system maps onto the
generic cavity opto-mechanical system perfectly. In this work we
approach the same problem from the perspective of diffraction of
matter-wave by an optical lattice. We will not restrict to the weak
excitation limit. We will develop a discrete-mode approximation
(DMA) for the condensate, and by comparing with the full description
of Gross-Pitaevskii (GP) equation, we find that the finite-mode
approximation is quantitatively good in the case of weak atom-atom
interaction. The DMA also enables us to find out the steady states
of the system, analyze their stability, and elucidate the
possibility of breakdown of adiabaticity and optical bistability in
this system.

This paper is organized as follows. The system and the general
formalism are described in Sec.~\ref{sec2}. Then in
Sec.~\ref{sec3a}, we compare the DMA with the full GP dynamics so as
to validate the DMA. In Sec.~\ref{sec3b}, we address the problem of
adiabatic evolution of the system, which is relevant to optical
bistability. Our results are summarized in Sec.~\ref{sec4}.

\section{general formalism}\label{sec2}
We assume that a cigar-shaped Bose-Einstein condensate is located
inside an optical cavity with its axial direction parallel to that
of the cavity. The internal transition frequency of the two-level
modeled atoms is $\omega_a$, while the relevant cavity mode
interacting with the atoms is of frequency $\omega_c$. The cavity
mode is coherently driven by a pump laser with frequency $\omega_p$
and a possibly time-dependent amplitude $\eta(t)$. The atom-pump and
cavity-pump detunings are denoted as $\Delta_a=\omega_a-\omega_p$
and $\Delta_c=\omega_c-\omega_p$, respectively. In the large
detuning limit and by transferring into the rotating frame at the
pump frequency, we get the Hamiltonian for the condensate-cavity
system ($\hbar=1$ throughout in this paper) \cite{esslinger_a},
\begin{eqnarray}
% \nonumber to remove numbering (before each equation)
  H =\int dx \hat{\Psi}^\dagger(x)\bigg( -\frac{1}{2m}\frac{d^2}{d^2 x}+V_{ext}(x)+U_0 \hat{a}^\dagger \hat{a} \cos^2(k
  x) &&\nonumber \\
   +\frac{g_{1D}}{2}\hat{\Psi}^\dagger(x)\hat{\Psi}(x)\bigg)
  \hat{\Psi}(x) \quad\quad\quad\quad\quad\quad\quad&& \nonumber \\
  +\Delta_c \hat{a}^\dagger \hat{a}
  +\eta(t)(\hat{a}+\hat{a}^\dagger)+H_{\kappa}. \quad\quad\quad\quad\quad&& \label{h1}
\end{eqnarray}
Here, $\hat{\Psi}^\dagger$ and $\hat{a}^\dagger$ are the creation
operators for the atoms (with mass $m$) and the cavity photons,
respectively. $V_{ext}(x)=\frac{1}{2} m \omega_{\parallel}^2 x^2$ is
the harmonic potential in the axial direction with characteristic
frequency $\omega_\parallel$, and $g_{1D}$ is the effective
atom-atom interaction strength in a transversely tight-confining
trap \cite{gorlitz}. Besides the harmonic potential $V_{ext}$ (which
may also owe to atom-photon interaction, as in an optical trap), the
atom-cavity photon interaction provides an additional potential $U_0
\hat{a}^\dagger \hat{a} \cos^2(k x)$ for the atoms. The difference
is that, the former is static, while the latter, being proportional
to the intra cavity photon number, may be dynamical and quantized
\cite{ritsh_prl_05}. Here $U_0=-g_0^2/\Delta_{a}$ is the maximal
light shift per photon an atom may experience (at an antinode), with
$g_0$ being the atom-photon coupling constant. Note that we assume
that the transverse radius of the condensate is much smaller than
the cavity mode waist, and therefore it is legitimate to neglect the
variation of $g_0$ in the transverse direction. The cavity mode
function in the axial direction is $\cos(k x)$, with the wave vector
$k=2\pi/\lambda$. Finally, $H_{\kappa}$ accounts for cavity photon
decay with a rate $\kappa$.

In the mean field approximation, we take $\hat{\Psi}(x,t) \sim
\Psi(x,t)$ and $\hat{a} \sim \alpha$, i.e., both the matter-wave
field and the electromagnetic field are described as classical
fields. The Gross-Pitaevskii (GP) equation for the condensate is
\begin{eqnarray}
% \nonumber to remove numbering (before each equation)
  i \frac{\partial \Psi(x,t)}{\partial t}  &=&\bigg( -\frac{1}{2m}\frac{d^2}{d^2 x}+V_{ext}(x)+U_0 |\alpha|^2 \cos^2(k
  x)   \nonumber\\
   && \quad  + g_{1D} |\Psi(x,t)|^2 \bigg)
   \Psi(x,t),   \label{condensate motion}
\end{eqnarray}
and the equation of motion for the cavity field is
%\begin{equation}\label{photon motion}
%\frac{\partial \alpha}{\partial t}=-i\left( \Delta_c +U_0 \int dx
%|\Psi(x,t)|^2 \cos^2(kx) -\kappa\right) \alpha +\eta(t).
%\end{equation}
\begin{eqnarray}
\frac{\partial \alpha}{\partial t}&=&-i\left( \Delta_c +U_0 \int dx
|\Psi(x,t)|^2 \cos^2(kx) \right) \alpha\nonumber \\
& &\quad\quad-\kappa \alpha +\eta(t). \label{photon motion}
\end{eqnarray}
We can numerically integrate these coupled equations and so as to
study the dynamics of the combined system. To this end, technically,
we had better convert the original equations into their
dimensionless form. The characteristic length scale and energy scale
of the system are $\xi=\lambda/2\pi$ and $\omega_r=k^2/2m$, i.e.,
the period of the optical lattice divided by $\pi$ and the atomic
recoil energy, respectively. We thus rescale position and time as
$x=\tilde{x} \xi$ and $t=\tilde{t}/\omega_r$ respectively, and
accordingly the condensate wavefunction $\Psi(x,t)=\sqrt{N/\xi}
\tilde{\Psi}(\tilde{x},\tilde{t})$, so that the scaled wavefunction
is normalized to unity, $\int d \tilde{x}|
\tilde{\Psi}(\tilde{x},\tilde{t})|^2=1$. Here $N$ is the total atom
number. Equations (\ref{condensate motion}) and (\ref{photon
motion}) then convert to
\begin{eqnarray}
% \nonumber to remove numbering (before each equation)
  i \frac{\partial \tilde{\Psi}(\tilde{x},\tilde{t})}{\partial \tilde{t}}  &=& \bigg( -\frac{d^2}{d^2 \tilde{x}}+\omega^2\tilde{x}^2+\tilde{U}_0 |\alpha|^2 \cos^2
  \tilde{x}   \nonumber\\
   & &\quad  + g |\tilde{\Psi}(\tilde{x},\tilde{t})|^2 \bigg)
   \tilde{\Psi}(\tilde{x},\tilde{t}), \label{bec motion 2}
\end{eqnarray}
%\begin{equation}\label{photon motion 2}
%\frac{\partial \alpha}{\partial \tilde{t}}=-i\big( \tilde{\Delta}_c
%+N \tilde{U}_0 \int d\tilde{x} |\tilde{\Psi}(\tilde{x},\tilde{t})|^2
%\cos^2\tilde{x} \big) \alpha-\tilde{\kappa} \alpha +\tilde{\eta}(t).
%\end{equation}
\begin{eqnarray}
% \nonumber to remove numbering (before each equation)
\frac{\partial \alpha}{\partial \tilde{t}}&=& -i\big(
\tilde{\Delta}_c +N \tilde{U}_0 \int d\tilde{x}
|\tilde{\Psi}(\tilde{x},\tilde{t})|^2 \cos^2\tilde{x} \big) \alpha
\nonumber \\
&&\quad\quad -\tilde{\kappa} \alpha +\tilde{\eta}(t).  \label{photon
motion 2}
\end{eqnarray}
Here we introduced the rescaled dimensionless quantities
$(\tilde{U}_0,\tilde{\Delta}_c,\tilde{\kappa},\tilde{\eta})=(U_0,\Delta_c,\kappa,\eta)/\omega_r$,
the rescaled harmonic frequency $\omega=\sqrt{m\omega_{\parallel}^2
\xi^2/2\omega_r}$, and the atom-atom interaction strength
$g=Ng_{1D}/\omega_r \xi$.

At this point, a further simplification is possible. Note that
experimentally the cavity damping is much faster than the mechanical
motion of the condensate. The former occurs at a rate of $\kappa$,
while the latter, roughly speaking, is on the order of $\omega_r$
(see the discrete-mode approximation below). In the experiment of
Esslinger \textsl{et al}. \cite{esslinger_a}, $\kappa=2\pi\times
1.3$ MHz, and $\omega_r$ is around $2\pi \times 3.75$ kHz
($^{87}$Rb), so the cavity decay is almost three orders of magnitude
faster than than the condensate motion. We thus can safely assume
that the cavity field follows the condensate adiabatically and solve
the photon number as
\begin{equation}\label{photon number}
  n_{ph}=|\alpha(t)|^2=\frac{\eta^2(t)}{\kappa^2 + (\Delta_c+NU_0 \langle \cos ^2
    \tilde{x}\rangle)^2},
\end{equation}
where $\langle \cos ^2 \tilde{x}\rangle$ is defined as $\int
d\tilde{x} |\tilde{\Psi}(\tilde{x},\tilde{t})|^2\cos^2\tilde{x} $,
and has the meaning of the overlap between the cavity mode intensity
and the condensate density distribution. Obviously, it is bounded
from above and below, $0\leq \langle \cos ^2 \tilde{x}\rangle \leq 1
$. Equation (\ref{photon number}) gives the dependence of the intra
cavity photon number, i.e., the intra cavity optical lattice depth
on the condensate. The cavity-pump detuning is shifted by the
condensate to an effective value $\Delta_{eff}=\Delta_c+NU_0 \langle
\cos ^2 \tilde{x}\rangle$.

Substituting Eq.~(\ref{photon number}) into Eq.~(\ref{bec motion
2}), we get a GP equation for the condensate, the most prominent
feature of which is that the potential acting on the condensate
depends in a highly non-local and non-linear way on the condensate
itself. This is to be compared with the non-linear term in the usual
GP equation, the atom-atom interaction, which is nonlinear but
local. If this cavity induced nonlinearity is to exhibit itself
apparently, the influence of the condensate on the cavity field
should be strong. To this end, it is desirable to have $\kappa
\lesssim NU_0$ (with an appropriate $\Delta_c$). Only under this
condition, the condensate is able to shift the cavity into or out of
resonance with the pump [see Eq.~(\ref{photon number})] and so as to
influence the cavity field drastically. It is worth noting that this
condition has been fulfilled with both Fabry-Perot cavities
\cite{esslinger_a,stamper-kurn_prl,stamper-kurn_NP} and ring
cavities \cite{hemmerich}.

Although a brutal integration of Eq.~(\ref{bec motion 2}) is
possible, and that captures all the possible atomic modes
simultaneously, we would like to introduce the discrete-mode
approximation (DMA). The advantage of DMA is that it captures the
physics and is technically simpler than the full GP equation. To
explain the basic idea, let us consider the extremal case $\omega=0$
and $g=0$, i.e., a non-interacting homogeneous condensate. The
ground state of the condensate is the zero momentum state $|p=0
\rangle$. The effect of the intra cavity optical lattice is to
diffract this state into the state $(|p=2k \rangle+ |p=-2k
\rangle)/\sqrt{2}$ (diffraction into the asymmetrical superposition
state is forbidden by the parity symmetry of the potential), which
can then be further diffracted into $(|p=4k\rangle+ |p=-4k
\rangle)/\sqrt{2}$ (or back into $|p=0 \rangle$), and so on. Thus
the dynamics of the system actually confines to some discrete atomic
modes. This stimulates us to introduce the DMA.

In the actual case, $\omega \neq 0$ and $g \neq 0$, the ground state
$\Psi_g(x)$ is not strictly the zero momentum state. However,
usually the length of the condensate is much larger than the period
of the optical lattice, thus by the long range coherence of the
condensate and the uncertainty relation, we infer that the momentum
distribution of the condensate is well localized around the origin
with a width much smaller than $2 k$. Thus the picture for the
non-interacting homogeneous BEC carries over to the actual case, as
long as in the period we are concerned with, the atomic diffraction
process overwhelms other effects caused by the finite trap potential
and atom-atom interaction. We can take into account the small but
finite momentum spread of the initial state by defining the discrete
modes as $\phi_0=\Psi_g(x)$, $\phi_1=\sqrt{2}\cos(2kx)\Psi_g(x)$,
$\phi_2=\sqrt{2}\cos(4kx)\Psi_g(x)$, and generally for $n\in
\mathbb{N}$, $\phi_n=\sqrt{2}\cos(2nkx)\Psi_g(x)$ \cite{implicit}.
In the momentum representation, $\phi_n$ is a coherent superposition
of two peaks localized at $\pm 2nk$ respectively. For the $n$-th
mode, we introduce the creation (annihilation) operator
$\hat{c}_n^\dagger$ ($\hat{c}_n$), with common boson commutation
relations. Note that these mode functions are not strictly
orthonormal, so this assignment of commutation relations is a bit
problematic. However, as long as the ground state $\Psi_g(x)$ varies
slowly on a length scale of $\lambda/2$, the error is negligible.

The DMA assumes that the condensate dynamics is confined in the
subspace spanned by $\{ \phi_n \}$. Numerically, we find that it is
enough to cut off at $\phi_2$. Higher modes play a minor role due to
the quadratically increasing eigenenergies. Substituting
$\Psi(x)=\sum_{i=0}^2 \hat{c}_i \phi_i(x)$ into Eq.~(\ref{h1}), and
neglecting the atom-atom interaction, we get the Hamiltonian in the
DMA
\begin{eqnarray}
% \nonumber to remove numbering (before each equation)
  H &=& 4\omega_r \hat{c}_1^\dagger \hat{c}_1 +16 \omega_r \hat{c}_2^\dagger \hat{c}_2 +\frac{U_0}{4}a^\dagger a \big(\sqrt{2}(\hat{c}_0^\dagger\hat{ c}_1 +\hat{c}_1^\dagger \hat{c}_0) \nonumber \\
   & & +\hat{c}_1^\dagger \hat{c}_2 +c_2^\dagger \hat{c}_1+2N \big)+\Delta_c a^\dagger
   a+\eta (a+a^\dagger)+H_{\kappa}. \label{h2}
\end{eqnarray}
In our calculations, we have used the equations:
\begin{subequations}
\begin{eqnarray}
% \nonumber to remove numbering (before each equation)
  \int dx \phi_0^*\left(  -\frac{1}{2m}\frac{d^2}{d^2 x}+V_{ext}\right)\phi_0\equiv \varepsilon_0,\quad\quad\quad\quad\quad\quad\quad&& \\
  \int dx \phi_m^*\left(  -\frac{1}{2m}\frac{d^2}{d^2 x}+V_{ext} \right)\phi_n \simeq \delta_{mn}(\varepsilon_0+4n^2 \omega_r),\quad&& \\
  \int dx \phi_m^*\cos^2(kx) \phi_n \simeq
  \frac{1}{4}(\delta_{m,n-1}+\delta_{m,n+1}),\quad\quad \quad\quad \quad&&
\end{eqnarray}
\end{subequations}
and dropped the constant term proportional to $\varepsilon_0$. These
equations become exact in the non-interacting homogenous case. In
the DMA Hamiltonian, the atom-photon interaction can be interpreted
in two ways. For the atoms, they are diffracted between adjacent
modes with coupling strengths proportional to the photon number; For
the photons, their frequency is renormalized proportional to the
interference of the atomic modes.

The atom-atom interaction we neglect contains terms like
$\hat{c}_1^\dagger \hat{c}_1 \hat{c}_0^\dagger \hat{c}_0$
\cite{aainter}, which shift the energies of the excited modes, and
more importantly, it mixes the discrete modes we retain and other
modes we discard in a four-wave mixing way. Therefore, the atom-atom
interaction impairs the discrete-mode approximation. However, since
the energy scale of the atom-atom interaction is on the order of the
chemical potential of the condensate $\mu=(3g\omega/4)^{2/3}$ (in
units of $\omega_r$, in the Thomas-Fermi regime), it may be
legitimate to neglect it in a time scale of $1/\omega_r$ if $\mu \ll
4$, the free frequency of the first excited mode in units of
$\omega_r$. In Esslinger \textsl{et al}.'s experiment, $4\omega_r
\simeq 2\pi \times 15$ kHz, while the chemical potential of the
condensate is about $ 2\pi \times 2.4 $ kHz.

From the Hamiltonian (\ref{h2}), by taking the mean field
approximation $\hat{c}_i\sim \sqrt{N}Z_i$, $a\sim\alpha$, and
integrating out the cavity field, we get the equation of motion for
the condensate
\begin{equation}\label{eom 2}
    i\frac{d }{d\tilde{t}}Z =H(n_{ph}) Z=\left(H_1+n_{ph} H_2\right)Z,
\end{equation}
with $Z=(Z_0,Z_1,Z_2)^T$, and
\begin{equation}\label{h1h2}
H_1=\left( \begin{array}{ccc} 0 & \quad & \quad \\ \quad & 4 & \quad
\\ \quad & \quad & 16 \end{array} \right ),\quad H_2=\frac{\tilde{U}_0}{4} \left( \begin{array}{ccc} 0 & \sqrt{2} & 0  \\ \sqrt{2} & 0 &
1
\\ 0 & 1 & 0 \end{array} \right ).
\end{equation}
The photon number is
\begin{equation}\label{pho dma}
   n_{ph}= |\alpha|^2= \frac{\tilde{\eta}^2}{\tilde{\kappa}^2 +(\tilde{\Delta}_c'+ 4N Z^\dagger H_2
    Z)^2},
\end{equation}
where $\tilde{\Delta}_c'=\tilde{\Delta}_c+N\tilde{U}_0/2$ is the
rescaled effective cavity-pump detuning with the condensate in its
ground state [$Z=(1,0,0)^T$]. We would also like to define the
general effective cavity-pump detuning
$\tilde{\Delta}_{eff}=\tilde{\Delta}_c'+ 4N  Z^\dagger H_2 Z$.
Equation (\ref{pho dma}) is just Eq.~(\ref{photon number}) in the
discrete-mode approximation.
\section{Application of the formalism}\label{sec3}
In this section, we apply the formalism developed in the preceding
section to two different cases. The first case deals with a constant
pump, which acts on the BEC-cavity system abruptly, the evolution of
the BEC-cavity system is studied. The second case considers the
scenario that the pump is turned on slowly and smoothly. We aim to
study the evolution of self-sustained steady state of the BEC-cavity
system as the parameters varied.
\subsection{Constant pump: validity of the discrete-mode approximation}\label{sec3a}
\begin{figure*}[thb]
\begin{minipage}[b]{0.45 \textwidth}
\includegraphics[width=\textwidth]{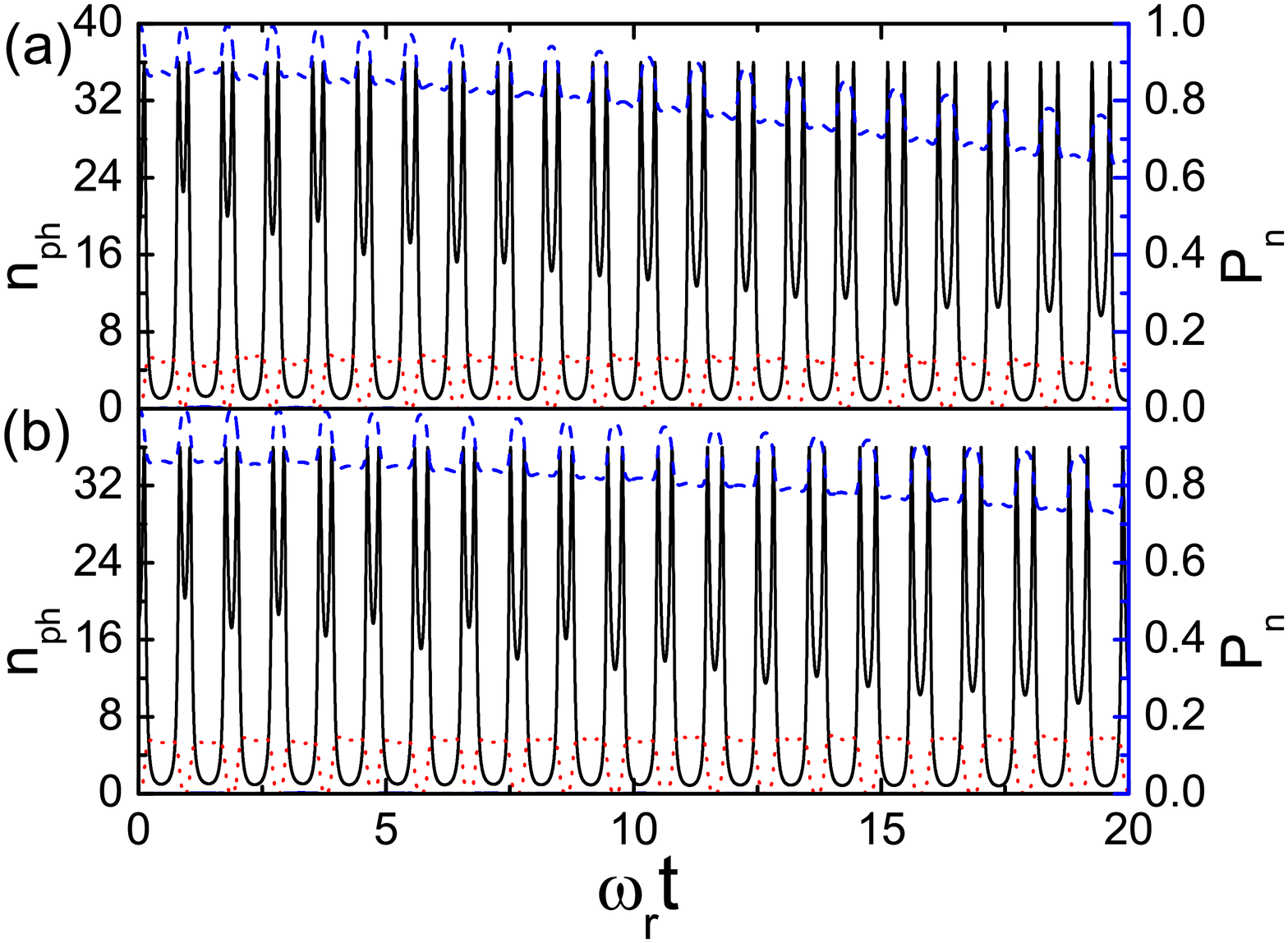}
\end{minipage}
\begin{minipage}[b]{0.45 \textwidth}
\includegraphics[width=\textwidth]{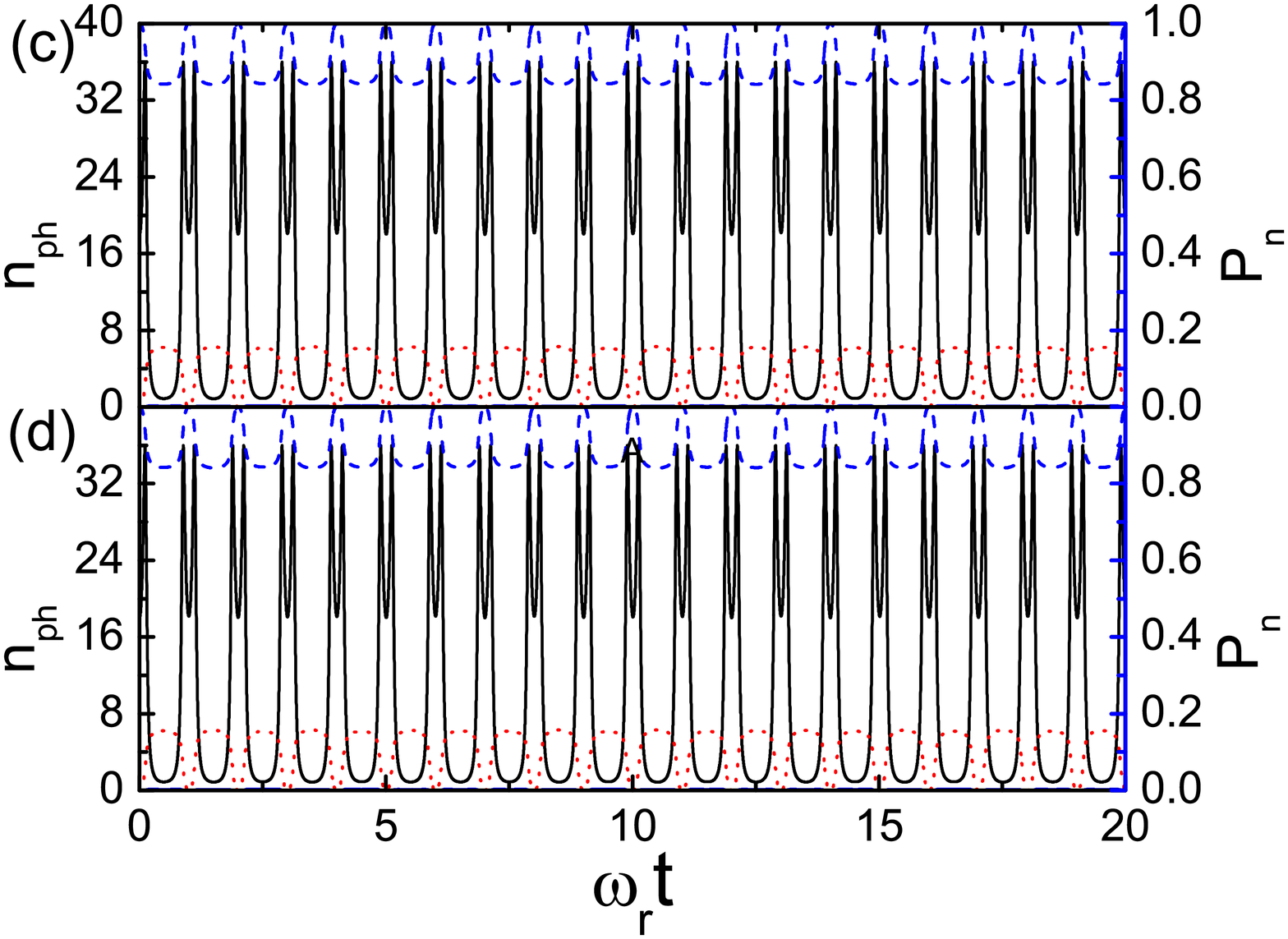}
\end{minipage}
\caption{\label{fig1}(Color online) Simulations of the BEC-cavity
dynamics. (a), (b) and (c): with Gross-Pitaevskii equation; (d):
with discrete-mode approximation. In each panel, the black solid
line is for the photon number, while blue dashed line, red dotted
line for the normalized populations on the atomic modes $\phi_0$ and
$\phi_1$, respectively (populations on the third mode are too small
and not shown). The parameters are $N=4.8\times 10^4$,
$\tilde{U}_0=0.25$,
$(\tilde{\kappa},\tilde{\Delta}_c',\tilde{\eta})=(0.4,0.4,2.4)\times
10^3$, $\omega=0.01$, and $g=100$, $50$, and $10$ in (a), (b) and
(c), respectively.}
\end{figure*}
\begin{figure}[thb]
\centering
\includegraphics[width=0.45\textwidth]{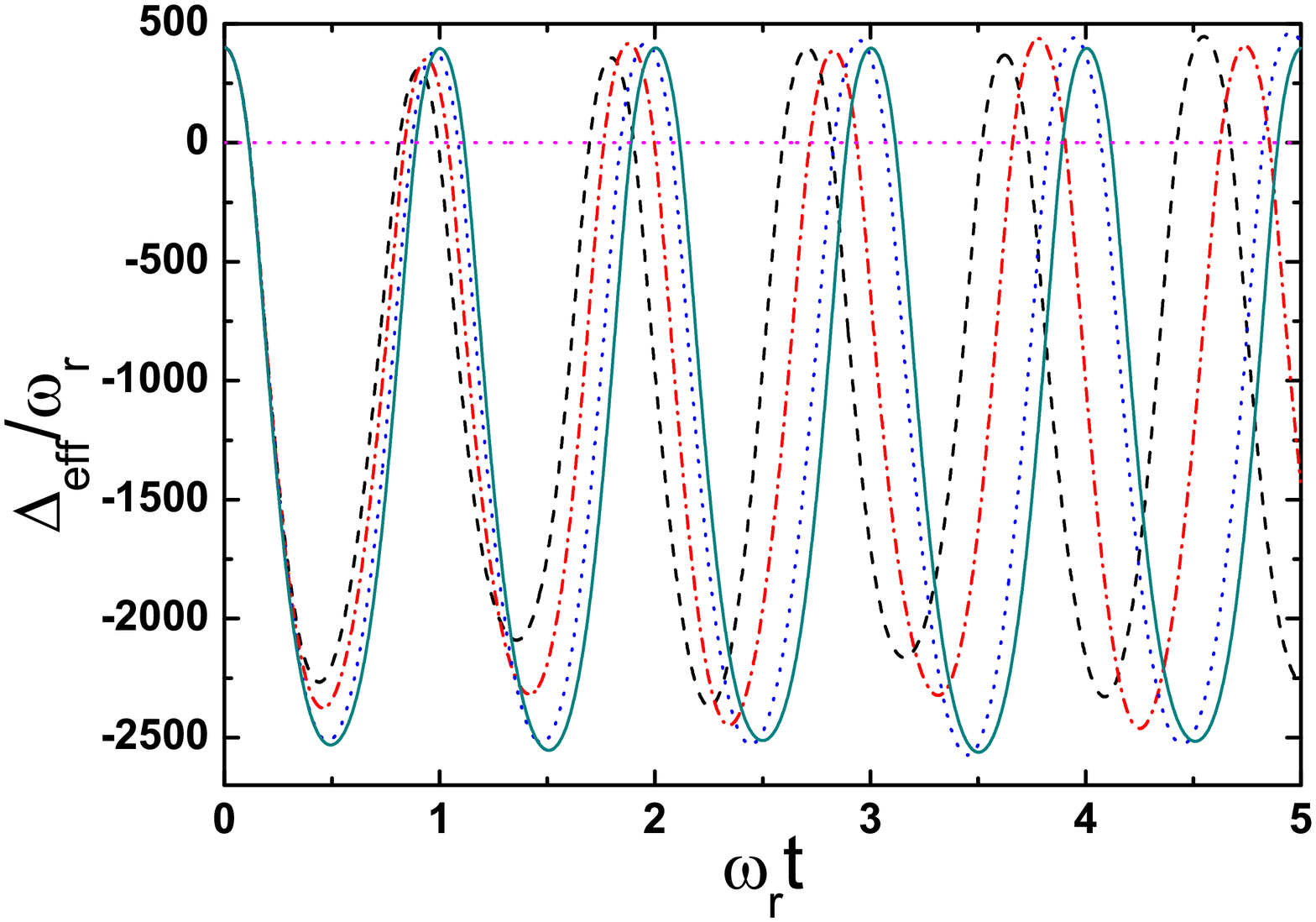}
\caption{\label{fig4}(Color online) Evolution of the effective
detuning $\Delta_{eff}$. The dashed line, dashed dotted line, dotted
line, and the solid line correspond to Fig.~\ref{fig1}(a)-(d),
respectively.}
\end{figure}
Assume that initially the BEC is in its ground state, and the cavity
is the vacuum. Then at $t=0$ a beam of laser with constant frequency
and constant amplitude is projected onto the cavity. In a time scale
of $1/\kappa$, the intra cavity field and hence the intra cavity
optical lattice builds up, which then diffracts the BEC into higher
momentum states. In the position representation, the condensate is
deformed by the optical lattice potential. The point is that, as the
condensate deforms, its overlap with the cavity mode and hence the
effective cavity-pump detuning varies, which in turn leads to the
variation of the lattice potential. Though atom diffraction by an
optical lattice has been extensively studied both theoretically and
experimentally since two decades ago
\cite{adams,phillips,pritchard,shore,kolovsky}, in all previous
works the optical lattice is either static or just a time-dependent
parameter. In contrast, here the lattice potential has to be treated
as a dynamical variable.

We have simulated the dynamics of the BEC-cavity system with both
the GP equation and the discrete-mode approximation. The GP
simulations are shown in Figs.~\ref{fig1}(a)-(c), with $\omega=0.01$
being fixed, and $g=100$, $50$, and $10$, respectively. In each
figure, we present the evolution of the photon number $n_{ph}$ and
the normalized populations $P_n=|\langle \phi_n|\Psi(t)\rangle|^2$
on different atomic modes. In one dimension, the Thomas-Fermi
approximation is valid if $g \gg \sqrt{2\pi \omega }$ \cite{tf}.
With this criterion, we find that the three cases are all deep in
the Thomas-Fermi regime. Thus, the atom-atom interaction plays an
important role in determining the ground state of the condensate. In
particular, the radius of the condensate is enlarged from
$(1/\omega)^{1/2}$ to $R_{TF}=(4g/3 \omega)^{1/3}$ (in units of
$\xi=\lambda/2\pi$). For all the three cases, we have $R_{TF}\gg
\pi$, therefore the basic condition for DMA is satisfied.

The simulation based on DMA is presented in Fig.~\ref{fig1}(d). Note
that in DMA, the atom-atom interaction and the initial ground state
are irrelevant. Thus, Fig.~\ref{fig1}(d) is a common approximation
to Figs.~\ref{fig1}(a)-(c). By comparing Fig.~\ref{fig1}(d) with
Figs.~\ref{fig1}(a)-(c), we see that to a time scale as long as
$\omega_rt=20$, the discrete-mode approximation agrees with the full
GP equation very well. Moreover, the agreement improves as the
atom-atom interaction decreases. The DMA recovers the $g=10$ case
[Fig.~\ref{fig1}(c)] almost exactly. By examining the difference of
Figs.~\ref{fig1}(a)-(c) with Fig.~\ref{fig1}(d), we identify two
consequences of the atom-atom interaction. Firstly, we see in
Figs.~\ref{fig1}(a) and \ref{fig1}(b) that as time goes on, the
total populations on the discrete modes we take decreases, and the
time scale of this process increases with decreasing atom-atom
interaction. Therefore, it is reasonable to attribute this
phenomenon to the atom-atom interaction, which populates the atoms
outside the modes we take. Secondly, the atom-atom interaction
shifts up the frequencies of the excited atomic modes, and thus
shortens the oscillation period of the system \cite{gorlitz}.

To illustrate the second point, we plot in Fig.~\ref{fig4} the
evolution of the effective detuning $\Delta_{eff}$ in
Figs.~\ref{fig1}(a)-(d). The oscillation of $\Delta_{eff}$ is
directly related to that of the photon number. We see clearly that
the periods in GP simulations are a bit shorter than that in DMA,
and the stronger the atom-atom interaction, the more apparent this
effect. Figure~\ref{fig4} also helps us understand the double-peak
structure of the photon number line in Fig.~\ref{fig1}. Each time
$\Delta_{eff}$ crosses zero, $n_{ph}$ attains its maximal value
$\tilde{\eta}^2/\tilde{\kappa}^2$.
\subsection{Slowly varying pump: breakdown of adiabaticity and optical bistability}\label{sec3b}
In this subsection, we proceed to consider the scenario that
initially the BEC is in its ground state and the cavity is in the
vacuum, and subsequently the pump is slowly and smoothly turned on.
In contrast to the situation above where the system is exposed
suddenly to a finite pump and experience oscillations indefinitely,
here we are concerned with the adiabatical following of the
self-sustained steady states. By a self-sustained steady state, we
mean that the two sides are stationary and consistent with each
other: the condensate is in the ground state determined by the
optical lattice (plus the harmonic trap), and meanwhile the optical
lattice (its depth) is determined by the condensate configuration,
i.e., fulfilling (\ref{photon number}) or (\ref{pho dma}). Naively,
it is expected that as long as the pump is ramped up sufficiently
slow, the BEC-cavity system will follow the steady states all the
way. However, now we are dealing with a nonlinear system and the
nonlinearity may give rise to the breakdown of adiabaticity
\cite{biao1,han,korsch,ling}. That is what we are interested in.

For our purpose, the key issue is to map out all the possible steady
states as the pump strength varies. Although this is a bit difficult
(time-consuming numerically) in the GP formalism, it can be done
very conveniently with DMA. In the framework of DMA, a steady state
corresponds to a solution of Eq.~(\ref{eom 2}) such that
\begin{equation}\label{steady 1}
    Z_0(\tilde{t})=Z_0 e^{-iE_0 \tilde{t}}.
\end{equation}
Or equivalently,
\begin{equation} \label{st2}
% \nonumber to remove numbering (before each equation)
  H(n_{ph})Z_0 = \left(H_1 +n_{ph} H_2\right)Z_0=E_0 Z_0,
\end{equation}
which defines a nonlinear eigenvalue problem because the Hamiltonian
$H$ depends on the eigenstate $Z_0$ through $n_{ph}$. It is
impossible to invoke the concepts and tools in linear algebra to
solve this problem. Our strategy is to first take an arbitrary trial
photon number $n_{tr}$, solve the ground state of the Hamiltonian
$H(n_{tr})$, $Z_{tr}$, and then substitute it into Eq.~(\ref{pho
dma}) to get an output photon number $n_{out}$. If $n_{out}=n_{tr}$,
then the solution is self-consistent and a steady state is obtained.
Note that $n_{out}$, as a function of $n_{tr}$, is continuous and
moreover, it is bounded both form above and below, $0 < n_{out} <
\tilde{\eta}^2/\tilde{\kappa}^2$. Therefore, if we scan $n_{tr}$
from $0$ to $\tilde{\eta}^2/\tilde{\kappa}^2$, by the principle of
continuity, there must be at least one point where $n_{out}=n_{tr}$.
This guarantees the existence of steady state solutions (clearly the
same strategy and arguments apply also in the GP formalism).
Furthermore, because of the nonlinearity of the system, the
possibility of more than one steady state solutions for a given set
of parameters is expected. As observed and argued in
Refs.~\cite{biao1,han,korsch}, the existence of more eigenvectors
than the dimensional of the Hilbert space is unique for nonlinear
systems.

\begin{figure*}[thb]
\begin{minipage}[b]{0.49 \textwidth}
\includegraphics[width=\textwidth]{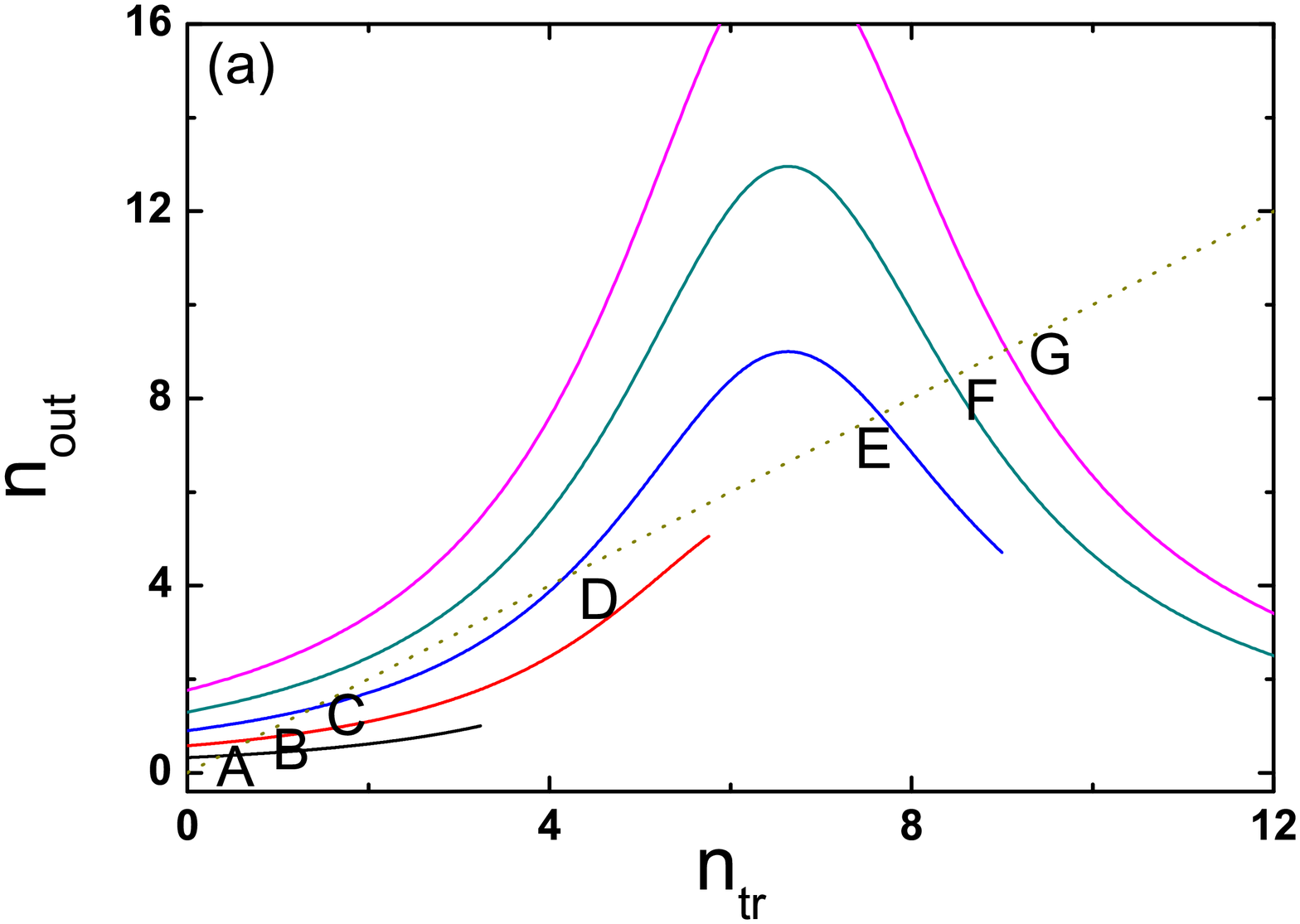}
\end{minipage}
\begin{minipage}[b]{0.45 \textwidth}
\includegraphics[width=\textwidth]{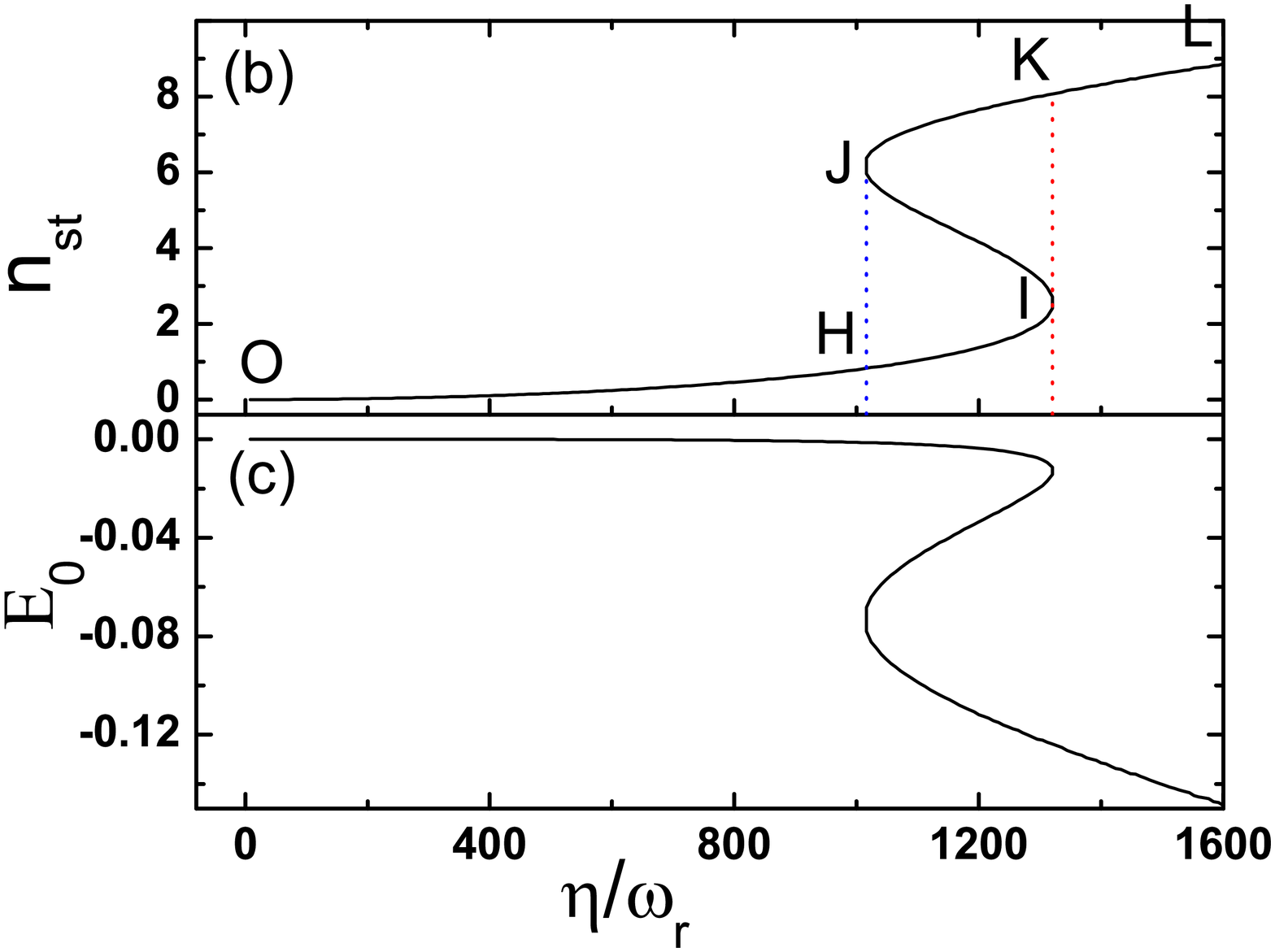}
\end{minipage}
\caption{\label{fig2}(Color online) (a) The output photon number
$n_{out}$ as a function of the trial photon number $n_{tr}$. The
intersection points ($A-G$) of the curves with the diagonal dotted
line correspond to steady states. The parameters are $N=4.8\times
10^4$, $\tilde{U}_0=0.25$, $(\tilde{\kappa},\tilde{\Delta}_c')=(0.4,
1.2)\times 10^3$, and from bottom to up $\tilde{\eta}=(0.72, 0.96,
1.20, 1.44, 1.68)\times 10^3$ respectively. (b) Photon number at
steady state $n_{st}$ and (c) nonlinear eigenvalue $E_0$ as a
function of the pump strength $\tilde{\eta}$. Parameters other than
$\tilde{\eta}$ are the same as in (a). The coordinates of $I$ and
$J$ are $(\tilde{\eta}_1,n_{st1})=(1296,2.416)$ and
$(\tilde{\eta}_2,n_{st2})=(1016,6.36)$, respectively.}
\end{figure*}

In Fig.~\ref{fig2}(a), we depict the output photon number $n_{out}$
as a function of the trial photon number $n_{tr}$ as the pump
strength $\tilde{\eta}$ varies while other parameters are held at
constant. For each curve, each of its intersections with the dotted
line $n_{out}=n_{tr}$ corresponds to a steady state. We see that for
$\tilde{\eta} \leq 960$ and $\tilde{\eta} \geq 1440$, there is only
one steady state solution, while in an intermediate range, there can
be three (for example, $\tilde{\eta}=1200$). This fact is made more
evident in Figs.~\ref{fig2}(b) and \ref{fig2}(c), where we plot the
photon number at steady state $n_{st}$ and the nonlinear eigenvalue
$E_0$ as a function of $\tilde{\eta}$ directly. In the interval
$(\tilde{\eta}_2,\tilde{\eta}_1)=(1016,1296)$, each $\tilde{\eta}$
corresponds to three steady states and hence three $n_{st}$'s and
three $E_0$'s. The folded level structure in Fig.~\ref{fig2}(c) is
similar to the looped level structure in Refs.~\cite{biao1,korsch}.

\begin{figure}[thb]
\centering
\includegraphics[width=0.45\textwidth]{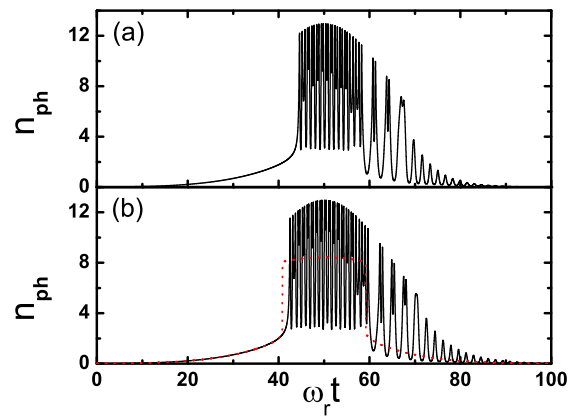}
\caption{\label{fig3}(Color online) Evolution of the cavity photon
number $n_{ph}$ as the pump is varied as
$\tilde{\eta}(t)=\tilde{\eta}_{max}\exp(-(\omega_r
t-50)^2/\sigma^2)$, with the peak amplitude
$\tilde{\eta}_{max}=1440$ and the pulse width $\sigma=31.25$, and
other parameters being the same as in Fig.~\ref{fig2}. (a)
simulation with GP equation ($\omega=0.01$ and $g=10$ as in
Fig.~\ref{fig1}(c)); (b) simulation with DMA, the dotted line
dictates the photon number of steady states drawn from
Fig.~\ref{fig2}(b). Note the sharp edges around $\omega_rt=40$.}
\end{figure}

With these figures as maps, we return to the scenario again. As the
pump is slowly ramped up from zero, the BEC-cavity system follows
the curve $OHI$ in Fig.~\ref{fig2}(b) adiabatically. Things go on
like this until the pump exceeds $\eta_{1}$, where the curve
terminates. Interpreted in terms of Fig.~\ref{fig2}(a), the system
first evolves from $A$ to $B$ to $C$, then at some point $C$
coincides with $D$ and then they disappear all together. Similar
interpretation can be made in terms of Fig.~\ref{fig2}(c). As $\eta$
crosses $\eta_{1}$, the system cannot jump discontinuously from $I$
to $K$. Therefore, beyond this critical value of pump, there is no
way for the BEC-cavity system to evolve adiabatically anymore, no
matter how slow the external pump is varied. We thus expect large
amplitude oscillations to appear abruptly as in Fig.~\ref{fig1}.
This is verified with simulations both based on the GP equation and
DMA, as shown in Fig.~\ref{fig3}. There we assume that the pump is
ramped up and down with a Gaussian profile. Both the simulations
reveal clearly a two-stage feature of the history of the BEC-cavity
system. Before a critical moment $\tilde{t}_c\sim 40$ (corresponding
to $\tilde{\eta}_c\sim \tilde{\eta}_{1}$), the photon number varies
smoothly and coincides with the prediction based on DMA perfectly.
This indicates the system evolves adiabatically along the line
$OHI$. However, just across this point, pronounced oscillations with
large amplitudes and small periods occur. Note that here the
breakdown of adiabaticity is related to the disappearance of a
nonlinear eigenstate. This is to be compared with its counterpart in
linear systems, which usually results from level-crossing. While the
latter can be avoided in principle, the former cannot. In the
viewpoint of state preparation, to prepare the system on the line
$KL$, we cannot take the route we take here in the parameter space.

The structure of the curve in Fig.~\ref{fig2}(b) reminds us of
optical bistability \cite{dispersive ob}. This motivates us to study
the dynamical stability of the steady states. We would like to
stress that adiabaticity and stability are two different but closely
related issues. Only those steady states which are dynamically
stable can be adiabatically evolved \cite{ling}. Let
$Z(\tilde{t})=Z_0(\tilde{t})+\delta Z(\tilde{t})$, with $\delta Z$
being an infinitesimal derivation from the solution $Z_0$.
Substituting this into the differential equation (\ref{eom 2}), and
making use of the fact $i\frac{d}{d\tilde{t}} Z_0=\left(H_1+H_2
n_{ph}(Z_0,\bar{Z}_0)\right)Z_0$, we get to the first order in
$\delta Z$,
\begin{eqnarray}
% \nonumber to remove numbering (before each equation)
  i\frac{d}{d\tilde{t}}\delta Z_i =\left(H_1^{ij}+  H_2^{ij}n_{ph}(Z_0,\bar{Z}_0) \right) \delta
  Z_j \quad \quad \quad \quad \quad \quad && \nonumber \\
    +H_2^{ij}Z_{0j}\left[\left(\frac{\partial
n_{ph}}{\partial Z_k}\right)_0\delta
   Z_k +\left(\frac{\partial n_{ph}}{\partial \bar{Z}_k}\right)_0\delta
   \bar{Z}_k\right].&&
\end{eqnarray}
Here the subscript $0$ means taking value at $(Z_0,\bar{Z}_0)$.
Assuming
\begin{equation}\label{delta z}
    \delta Z(t)=e^{-iE_0 t}\left (ue^{-i\Omega t}-v^* e^{i \Omega t}
    \right),
\end{equation}
and substituting this into the equation above, we get
\begin{equation}\label{matrix m 1}
 M \left (\begin{array}{c} u\\v \end{array}\right) =\Omega \left( \begin{array}{c} u\\v
    \end{array} \right),
\end{equation}
where
\begin{equation}\label{sigma and m}
 M=\left (\begin{array}{cc} A+B-E_0 & -B \\ B & -A-B+E_0
    \end{array}\right).
\end{equation}
Here the $3 \times 3$ matrices $A$ and $B$ are defined as
\begin{eqnarray}
% \nonumber to remove numbering (before each equation)
  A &=& H_1 + n_{ph}(Z_0,\bar{Z}_0) H_2,\nonumber \\
  B_{ij} &=& \left( H_2^{ik} Z_{0k}\right) \left(\frac{\partial n_{ph}}{\partial Z_j}
  \right)_0,\quad 0 \leq i,j,k \leq 2.
\end{eqnarray}
The dynamical stability of the steady states is determined by the
eigenvalues of the matrix $M$. If all the eigenvalues are real, the
steady state is dynamically stable; otherwise, it is dynamically
unstable.

\begin{figure}[thb]
\centering
\includegraphics[width=0.45\textwidth]{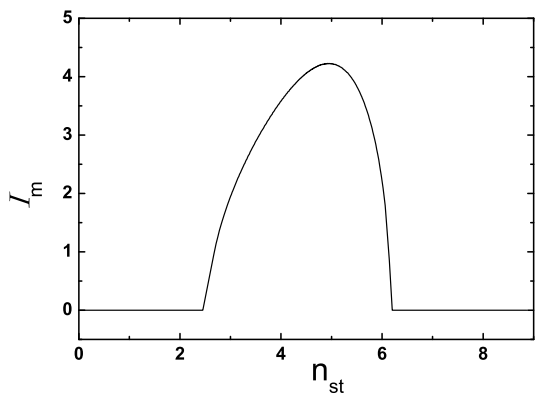}
\caption{\label{fig5} Maximal imaginary part $I_m=\max_i\{
Im(\lambda_i)\}$ of the eigenvalues of $M$. The points on the curve
$OHIJKL$ in Fig.~\ref{fig2}(b), which is parameterized with
$n_{st}$, are examined. The interval where $I_m$ is nonzero is
$(n_{st1},n_{st2})$.}
\end{figure}

In Fig.~\ref{fig5}, we examine the stability of the steady states on
the curve $OHIJKL$ by plotting the maximal imaginary part
$I_m=\max_i\{ Im(\lambda_i)\}$ of the eigenvalues of $M$ (as a real
matrix, its eigenvalues appear in conjugate pairs). The curve is
parameterized with the photon number $n_{st}$. We see that for
$n_{st}< n_{st1}$ ($n_{st} > n_{st2}$), which corresponds to $OHI$
($JKL$), the eigenvalues are all real and thus steady states on
these segments are stable. While for $n_{st1} < n_{st} < n_{st2}$,
which corresponds to $IJ$, some eigenvalues are complex and thus
steady states on this segment are unstable. This is consistent with
the usual observation that the negative sloped parts are always
unstable though the positive sloped parts are not necessarily stable
\cite{dispersive ob,ikeda}. As noted previously, stability is a
prerequisite of adiabaticity. Here, it is implied that we can
propagate adiabatically the BEC-cavity system along the line $OHI$
(as we do above), but it is not feasible to do that along $IJ$.

The analysis of stability confirms the possibility of optical
bistability in the BEC-cavity system. In previous experiments
dealing with a cloud of thermal cold atom gas
\cite{stamper-kurn_prl}, optical bistability has been discussed and
observed, but an analysis of stability is lacking. If there are some
dissipation channels, it may be possible to switch between the upper
and lower stable branches and observe the typical hysteresis loop of
optical bistability in the BEC-cavity system. Here we can only
simulate the coherent behavior of the system and thus this
phenomenon is beyond our means. However, the mechanism underlying
optical bistability still manifests itself in the form of breakdown
of adiabaticity.
\section{conclusions}\label{sec4}
We have investigated the nonlinear dynamics of a BEC-cavity system.
This system is characterized by the nonlinearity that the optical
lattice acting on the condensate depends on the condensate itself in
a highly non-linear and non-local way. By borrowing ideas from atom
optics, we develop the discrete-mode approximation (DMA). The
philosophy of DMA is that the dynamics of the condensate confines to
some discrete standing-wave modes. It is numerically verified that
in the limit of weak atom-atom interaction, DMA agrees with the full
GP equation dynamics perfectly. Our emphasis is on the adiabatical
evolution of the BEC-cavity system. The DMA facilitates us to solve
the nonlinear eigenstates of the system and analyze their stability.
Due to the nonlinearity of the system, the nonlinear energy level
can be folded in some parameter regimes. This structure gives rise
to both the breakdown of adiabaticity and optical bistability in
this system. As an open problem, we conjecture that these two
effects are complementary to each other. If the dynamics of the
system is purely coherent, the breakdown of adiabaticity will show
up; otherwise, if dissipations such as atom loss are strong, optical
bistability may take the place.

\section{acknowledgments}\label{sec5}
This work was supported by NSF of China under Grant Nos.~10874235,
60525417, and 10775176, NKBRSF of China under Grant
Nos.~2009CB930704, 2006CB921400, 2006CB921206, and 2006AA06Z104. J.
M. Z. would like to thank J. Ye for stimulating discussions.


\begin{thebibliography}{99}
\bibitem{implicit}
The atom-atom interaction is taken into account implicitly in the
definition of atomic modes, but not in the dynamics.
\bibitem{aainter}
Terms like $c_1^\dagger c_1^\dagger c_0 c_0$ are energetically
detuned, and thus play a relatively minor role if the chemical
potential of the condensate is much smaller than the energy
detuning, i.e., $\mu=(3g\omega/4)^{2/3} \ll 4$.
\end{thebibliography}
\end{document}